\documentclass[twocolumn]{article}

\usepackage[switch]{lineno}

\usepackage{graphicx}
\usepackage{titling}
\usepackage{microtype}
\usepackage[sc]{mathpazo}

\newcommand{\aap}{Astronomy and Astrophysics}

\newcommand{\aj}{Astronomical Journal}
\newcommand{\apj}{Astrophysical Journal}
\newcommand{\apjl}{Astrophysical Journal Letters}
\newcommand{\apjs}{Astrophysical Journal Supplement}
\newcommand{\araa}{Annual Review of Astronomy and Astrophysics}
\newcommand{\grl}{Geophysics Research Letters}

\newcommand{\icarus}{Icarus}
\newcommand{\jgr}{Journal of Geophysics Research}
\newcommand{\mnras}{Monthly Notices of the Royal Astronomical Society}
\newcommand{\nat}{Nature}
\newcommand{\pasp}{Publications of the Astronomical Society of the Pacific}

\newcommand{\psj}{Planetary Science Journal}

\newcommand{\ssr}{Space Science Reviews}

\pretitle{\begin{center}\Large\bf}
\posttitle{\end{center}}

\title{Atmospheric Dynamics of a Near Tidally Locked Earth-Size
  Planet}

\author{
\bf{Stephen R. Kane} \\
\normalsize Department of Earth and Planetary Sciences, University of
  California, Riverside, CA 92521, USA \\
\normalsize skane@ucr.edu 
}

\date{}

\renewcommand{\maketitlehookd}{%
\begin{abstract}

The discovery and characterization of Earth-sized planets that are in,
or near, a tidally-locked state are of crucial importance to
understanding terrestrial planet evolution, and for which Venus is a
clear analog. Exoplanetary science lies at the threshold of
characterizing hundreds of terrestrial planetary atmospheres, thereby
providing a statistical sample far greater than the limited inventory
of terrestrial planetary atmospheres within the Solar System. However,
the model-based approach for characterizing exoplanet atmospheres
relies on Solar System data, resulting in our limited inventory being
both foundational and critical atmospheric laboratories. Present
terrestrial exoplanet demographics are heavily biased toward
short-period planets, many of which are expected to be tidally locked,
and also potentially runaway greenhouse candidates, similar to
Venus. Here we describe the rise in the terrestrial exoplanet
population and the study of tidal locking on climate
simulations. These exoplanet studies are placed within the context of
Venus, a local example of an Earth-sized, asynchronous rotator that is
near the tidal locking limit. We describe the recent lessons learned
regarding the dynamics of the Venusian atmosphere and how those
lessons pertain to the evolution of our sibling planet. We discuss the
implications of these lessons for exoplanet atmospheres, and outline
the need for a full characterization of the Venusian climate in order
to achieve a full and robust interpretation of terrestrial planetary
atmospheres.

\end{abstract}

}

\begin{document}

\maketitle


\section{Introduction}
\label{intro}

Exoplanetary science has undergone rapid expansion over the past three
decades. As observational techniques have improved, the sensitivity
limits have allowed the detection of increasingly smaller exoplanets
\cite{butler2006,akeson2013}. The ``deep dive'' into the terrestrial
regime of exoplanet detection was largely enabled by the {\it Kepler}
mission \cite{borucki2016}; a legacy that is now being continued with
the Transiting Exoplanet Survey Satellite ({\it TESS})
\cite{ricker2015}. Moreover, these discoveries have revealed the
diversity of planetary system architectures \cite{ford2014,winn2015}
including compact systems of planets close to the host star
\cite{funk2010,kane2013e}. The proximity of exoplanets to their host
star has spurred further study into tidal locking scenarios, including
the time scales and implications for the planets
\cite{barnes2017b}. As such, the local analogs to tidally locked
terrestrial planets are exceptionally valuable subjects of
examination, as they provide the means to study the pathway to tidal
locking and the influence on their atmospheric evolution.

A particular area of focus has been the investigation of planets that
reside within their star's Habitable Zone (HZ), defined as the region
around a star where a planet may have surface liquid water provided
sufficient atmospheric pressure
\cite{kasting1993a,kane2012a,kopparapu2013a,kopparapu2014}. Many HZ
planets have been detected, including several hundreds from the {\it
  Kepler} mission \cite{kane2016c}. In parallel, the development of
climate models and their applications to exoplanets has seen a similar
rise in activity \cite{forget2014,shields2019a}. These climate
simulations are often constructed from parent Earth-based models
\cite{way2017b}, and require a substantial amount of assumptions
regarding intrinsic planetary properties
\cite{fauchez2021}. Furthermore, the effect of tidal locking on
potential surface habitability is unresolved, and may accelerate
\cite{hamano2013,turbet2021} or decelerate \cite{way2016} the
transition of terrestrial planet atmospheres into a runaway greenhouse
state. It is expected that exoplanet atmospheric compositional data
will provide additional insights that may substantially aid the
climate modeling approach \cite{kempton2018}. Importantly, the
detailed inference of planetary surface conditions relies upon a
complete understanding of the limited inventory of terrestrial
atmospheres within the Solar System, for which in-situ data form the
basis of atmospheric models \cite{kane2021d}.

The nearest Earth-size planet is Venus, whose complicated atmosphere
remains an enigma in many respects \cite{taylor2009,taylor2018}. The
surface conditions are among the most extreme observed in the Solar
System, with an atmospheric mass two orders of magnitude larger than
Earth, creating a surface pressure that renders its CO$_2$ dominated
atmosphere in a super-critical fluid state
\cite{lebonnois2020a}. Diagnosing the evolution of Venus from
formation to its present state is fundamental to understanding the
divergence of Venus and Earth \cite{kasting1988c} and possible similar
evolutionary pathways for exoplanets \cite{kane2019d}. Venus is
considered an asynchronous rotator \cite{leconte2015a}, whose present
rotational state is likely the result of complex interactions between
solar and atmospheric tides
\cite{ingersoll1978b,correia2001,auclairdesrotour2017b}. Similarly,
the dynamics and super-rotation of the Venusian atmosphere is a
consequence of the combined effects of relatively high incident flux,
slow rotation, atmospheric structure, and interaction with the surface
topography \cite{lee2007c,takagi2007,lebonnois2010}. The particular
relevance to exoplanets that lie near the tidal locking threshold is
acute \cite{sergeev2020}, for which Venus will always be our best
studied example \cite{kane2019d}.


\section{The Terrestrial Exoplanet Population}
\label{population}

\begin{figure}
  \begin{center}
      \includegraphics[width=0.9\linewidth]{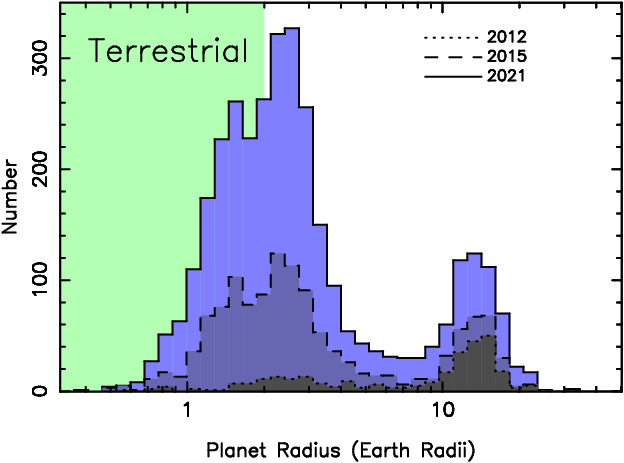}
  \end{center}
  \caption{Histogram of the radius distribution for the known
    exoplanet population. The largest histogram (solid line, light
    blue) represents the currently known distribution. The middle
    histogram (dashed line, medium blue) and the smallest histogram
    (dotted line, dark blue) represent the distributions as of 2015
    and 2012, respectively. The green shaded region indicates which
    planets may be terrestrial in nature.}
  \label{fig:radii}
\end{figure}

The currently known exoplanet inventory consists of a population
number that lies in the thousands \cite{akeson2013}, from which the
diversity of system architectures begins to emerge
\cite{ford2014,clanton2014b,winn2015}. Shown in Figure~\ref{fig:radii}
are histograms of the exoplanet radii distribution for discoveries up
until 2012, 2015, and 2021. The data were extracted from the NASA
Exoplanet Archive \cite{akeson2013} and are current as of 2021, June
10. The green shaded region indicates which planets may be
terrestrial, accounting for uncertainties in the radii measurements
\cite{rogers2015a,wolfgang2016}. The figure demonstrates the dramatic
rise in terrestrial planet detection that occurred between 2012 and
2021, largely due to the results from the {\it Kepler} mission
\cite{borucki2016}. Figure~\ref{fig:radii} also shows the gradual
appearance of the a gap in the radius distribution, referred to as the
``Fulton gap'' or ``photoevaporation valley''. This radius gap has
been suggested to be the result of photo-evaporation
\cite{lopez2013,owen2013a,fulton2017}, suggesting that short-period
super-Earths may primarily consist of a ``bare'' core from which the
atmosphere has been stripped within 100~Myrs of formation. Though the
transit and radial velocity exoplanet detection techniques are biased
toward short-period planets, the occurrence of terrestrial planets
within their sensitivity regimes have been broadly quantified
\cite{dressing2013,bryson2021}. These occurrence rates include those
for planets that lie interior to the system HZ, where the planets may
be stronger candidates for atmospheric evolution into post runaway
greenhouse environments \cite{kane2014e}.

\begin{figure*}[t]
  \begin{center}
      \includegraphics[width=0.9\linewidth]{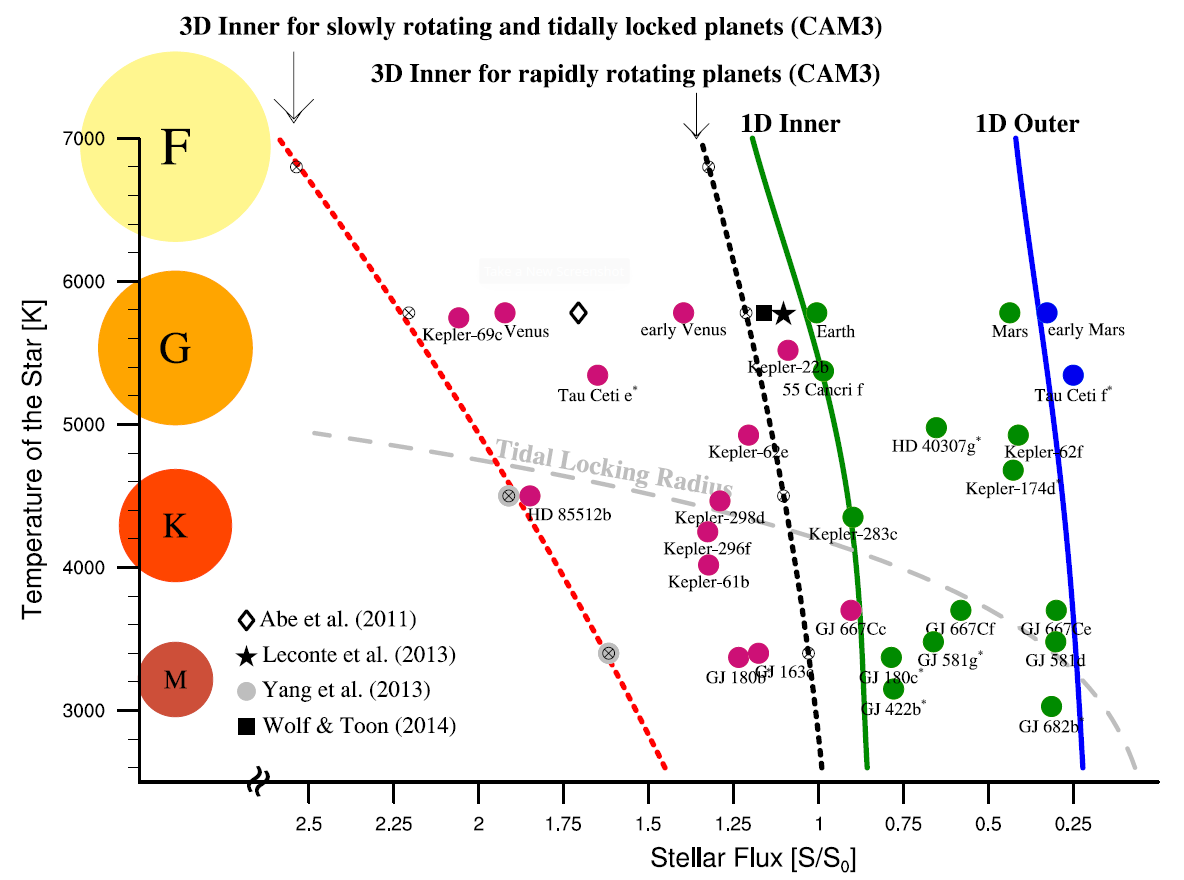}
  \end{center}
  \caption{The HZ boundaries (solid and dotted lines) and tidal
    locking distance (gray dashed line) as a function of stellar
    effective temperature and stellar flux received by the
    planet. Solar System planets and numerous known exoplanets are
    included \cite{yang2014b}.}
  \label{fig:tidallock}
\end{figure*}

The vast majority of detected exoplanets are relatively close to their
host stars, and are often considered to be susceptible to tidal
locking \cite{barnes2017b}, a process that can have significant
consequences for climate evolution
\cite{barnes2013a,yang2014b,leconte2015a,kane2020e}. Furthermore, the
rotational state of a planet can be a strong driver of its magnetic
field strength \cite{christensen2006,driscoll2015} which, in turn, may
play an important role in atmospheric erosion
\cite{zhang2009,gunell2018a}. Shown in Figure~\ref{fig:tidallock} is a
representation of HZ boundaries (solid and dotted lines) and tidal
locking distance (gray dashed line) as a function of stellar effective
temperature and stellar flux received by the planet
\cite{yang2014b}. The figure also includes numerous known exoplanets,
where those below the gray dashed line are considered to be tidally
locked to the host star. Note that those planets near or below the
tidal locking boundary are dominated by systems with Gliese host stars
which, since the catalog only extends to 25~pcs, tend to be low-mass
stars \cite{stauffer2010b}. The dotted lines are modified HZ
boundaries depending on the rotational state of the planet. It is
worth noting that there are many other aspects of the planet and star,
such as planetary obliquity and stellar type, that play major roles in
governing the atmospheric evolution of terrestrial planets.

The present state of a planet's angular momentum depends on the age of
the system and the tidal forces that are acting upon the planet. The
time scale for a planet to become tidally locked to the host star may
be expressed as
\begin{equation}
  \tau_\mathrm{lock} = \frac{\omega a^6 I Q}{3 G M_\star^2 k_2 R_p^5}
  \label{eq:lock}
\end{equation}
where $\omega$ is the rotation rate, $a$ is the semi-major axis, $I$
is the moment of inertia, $Q$ is the dissipation function, $G$ is the
gravitational constant, $M_\star$ is the mass of the star, $k_2$ is
the Love number, and $R_p$ is the radius of the planet
\cite{gladman1996b}. Note that, with the exception of the stellar
mass, all variables in Equation~\ref{eq:lock} pertain to the planetary
properties. The known properties of Venus may be used to thus estimate
the time until Venus may be tidally locked. Based on the analysis of
{\it Magellan} and {\it Pioneer Venus} data, we adopt values of $Q =
12$ and $k_2 = 0.299$ \cite{konopliv1996,dumoulin2017}. The subsequent
calculations yield a time scale for Venus tidal locking of $\sim 6.5
\times 10^6$~years. Such a relatively short time period must account
for the highly uncertain values of $Q$ and $k_2$. Additionally, the
precise nature of the change in the Venusian angular momentum remains
unclear \cite{bills2005e,campbell2019}, and the above calculation
neglects the non-negligible interaction of the atmosphere with the
surface \cite{leconte2015a,auclairdesrotour2017b}. Therefore, the
climate dynamics of a planet such as Venus are critical to
understanding the rotational evolution for those planets susceptible
to tidal locking.


\section{Climates of Tidally Locked Worlds}
\label{climates}

\begin{figure*}[t]
  \begin{center}
      \includegraphics[width=0.9\linewidth]{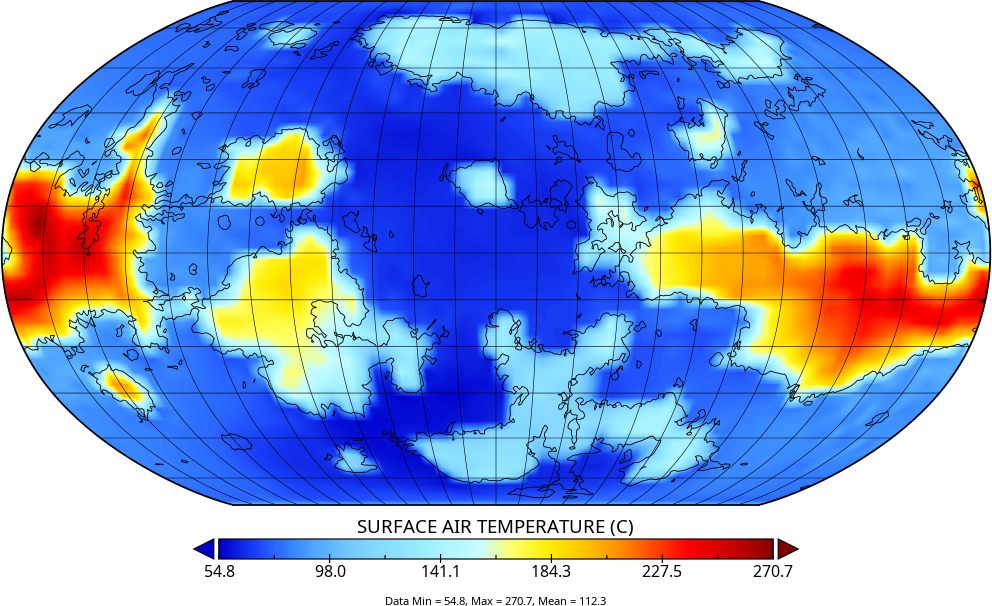}
  \end{center}
  \caption{A Robinson projection of the results from a ROCKE-3D
    simulation for the exoVenus candidate, Kepler-1649b
    \cite{kane2018d}, showing the surface temperature and a Venus
    topography overlay. The surface temperature reached a maximum of
    270$^\circ$C and a global mean of 112$^\circ$C at the limit of the
    ROCKE-3D radiation tables.}
  \label{fig:rocke3d}
\end{figure*}

As described in Section~\ref{population}, a significant fraction of
the known terrestrial exoplanet population are expected to be near or
in a tidally locked state. Consequently, significant effort has been
expended toward adapting climate models to slow rotational states,
including synchronized rotation. Functionally, this is generally
achieved by setting the rotation period to a value similar to the
orbital period, and adjusting the spectral energy distribution
received at the top of the atmosphere. The application of these
simulations has been successfully demonstrated on numerous occasions,
some with far-reaching consequences. For example, climate simulations
of tidally locked planets have revealed conditions that lead to
climate instability, whereby positive or negative feedbacks result in
a runaway climate shift \cite{kite2011c}, and potential atmospheric
collapse, whereby cooling and condensation of the atmosphere result in
dramatic changes in atmospheric conditions
\cite{wordsworth2015a,auclairdesrotour2020}. Others have studied the
relationship between atmospheric circulation, convection, and thermal
structure that tidal locking produces
\cite{koll2016,sergeev2020,hammond2021}. The HZ boundaries,
particularly the inner edge, have been found to have a strong
rotational dependence, emphasizing the importance of tidal locking for
target selection of potentially habitable worlds
\cite{yang2013,leconte2015a}. Additional simulations have shown the
effects of water trapping and specific cloud features on the dayside
of tidally locked planets
\cite{yang2014c,yang2019b,ding2020a}. Climate simulations have also
demonstrated the dependence of the climate dynamics of tidally locked
planets on surface topography \cite{carone2016,lewis2018}, discussed
further in Section~\ref{venus}.

To date, climate simulations have been applied to numerous known
exoplanets suspected of being in a tidally locked state, including the
nearest exoplanet: Proxima Centauri b
\cite{turbet2016,delgenio2019a}. An example of such simulations is the
case of Kepler-1649b, a short-period planet orbiting an M dwarf star
that is a strong candidate as an exoVenus analog
\cite{angelo2017a}. Various simulations of the climate evolution using
the general circulation model ROCKE-3D \cite{way2017b} found that, in
all cases, the surface temperature quickly rises to the limit of the
radiation tables utilized, indicating a progression into a moist or
runaway greenhouse state \cite{kane2018d}. Shown in
Figure~\ref{fig:rocke3d} is an example output from these simulations
in the form of a Robinson projection of the final surface temperature
map with a Venus topography overlay. The initial conditions in this
case included an incident flux of double the solar constant, CO$_2$
and CH$_4$ atmospheric compositions of 400 and 1 ppmv, respectively,
and a paleo-Venus topography with all-ocean grid cells set to 1360~m
in depth. Before reaching the limit of the radiation tables, the
surface temperature reached a maximum value of 270$^\circ$C, with a
global mean of 112$^\circ$C. Similar simulations have been used to
suggest that Venus may have experienced a significant period of
temperate surface conditions \cite{way2016}, with substantial cloud
feedback near the substellar point producing a cooling effect
\cite{way2020}. Such scenarios assume that water inventory in the
atmosphere was able to condense at the surface to form oceans, which
may have been prevented via early magma oceans and/or insolation flux
\cite{hamano2013,turbet2021}. Regardless, the atmospheric dynamics of
a nearby, slowly rotating, Earth-size planet, and how it has evolved
through time, is a critically important laboratory to test many of
these principles.


\section{Venus Atmospheric Dynamics}
\label{venus}

Despite a slow rotation, Venus maintains an extremely dynamic
climate. The Venusian surface conditions are exceptionally different
from those found on Earth, with a mean surface temperature and
pressure of $\sim$735~K and 93~bar, respectively. Although these
surface conditions are globally consistent, the day-side and
night-side temperature diverge considerably above an altitude of
$\sim$100~km \cite{schubert1980c}. Since only $\sim$3\% of the
sunlight incident at the top of the atmosphere reaches the surface,
almost all of the solar energy absorbed by the planet is deposited
into the atmosphere \cite{crisp1986a}. Despite the slow rotation and
nearly zero eccentricity and obliquity, the atmosphere of Venus
consistently produces convective cells that results in a non-uniform
atmospheric circulation \cite{schubert1980c}. Furthermore, the
turbulence produced in the atmosphere may be the driver for the
super-rotation of the middle and upper atmosphere
\cite{horinouchi2020}, which rotates 60 times faster than the solid
planet, and may play an important role in the dynamics of many
exoplanetary atmospheres \cite{imamura2020}. The super-rotation for
terrestrial planets differs from those that occur for giant planets,
such as Jupiter and Saturn, due to the interaction of terrestrial
atmospheres with the solid planet and the internal heat flux and
interior convection of giant planets that provides an additional
driver of super-rotation in the cloud top layers \cite{kaspi2020}. The
atmospheric super-rotation observed for Venus (and also Titan) and the
mechanisms that drive them thus remain a significant challenge for
numerical models \cite{read2018b}.

\begin{figure}
  \begin{center}
      \includegraphics[width=0.9\linewidth]{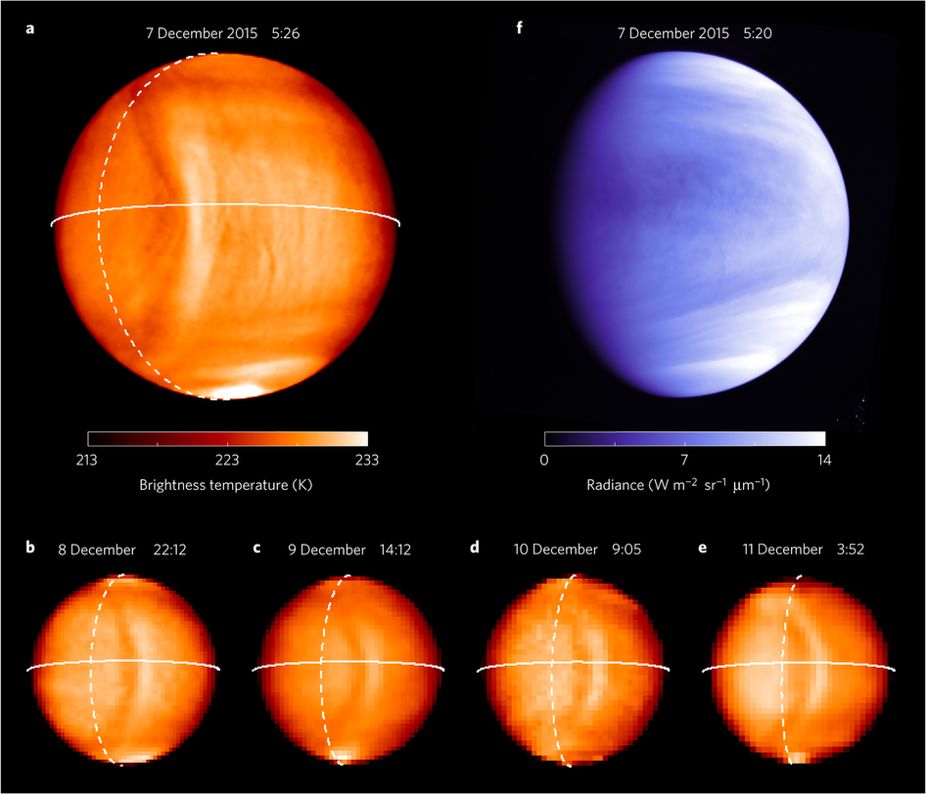}
  \end{center}
  \caption{{\it Akatsuki} data showing the brightness temperature
    (a--e) and UV brightness (f) of the Venusian disk. The bow-shaped
    structure is an atmospheric gravity wave passing over the surface
    topography \cite{fukuhara2017a}.}
  \label{fig:akatsuki}
\end{figure}

Climate simulations of tidally locked planets have produced strong
evidence that climates can interact significantly with surface
topography \cite{carone2016,lewis2018}. Such evidence has been
observed directly in the Venusian atmosphere, such as the detection
and subsequent analysis of stationary gravity waves detected using
data from the {\it VEGA} Balloon \cite{young1987b,young1994c}, and the
{\it Venus Express} mission \cite{peralta2017c}. These observations
were verified by data from the {\it Akatsuki} mission, which indicate
the presence of stationary waves in the upper atmosphere of Venus
\cite{fukuhara2017a}, as shown in Figure~\ref{fig:akatsuki}. The
center of the wave features, visible in the brightness temperature
maps (a--e), are located approximately above the Aphrodite Terra
highland region, which suggest that the causation is the direct result
of deep atmosphere movement over, and interaction with, the elevated
terrain \cite{bertaux2016}. The structure of these waves is not
uniform, indicating a dependence on latitudinal and diurnal effects,
and thus their production resulting from a complex interaction between
atmospheric dynamics and solar heating \cite{kouyama2017}. The
explanation of the atmospheric features as stationary gravity waves
was further validated via numerical simualtions that also contribute
to the atmospheric torque that acts upon the solid planet
\cite{navarro2018}. This interplay between incident flux, rotation,
atmospheric dynamics, and topography leads to correlated behavior that
is both important to include in simulations and may also enable
inferences regarding surface conditions and topography.


\section{Observational Diagnostics}
\label{model}

\begin{figure}
  \begin{center}
      \includegraphics[width=0.9\linewidth]{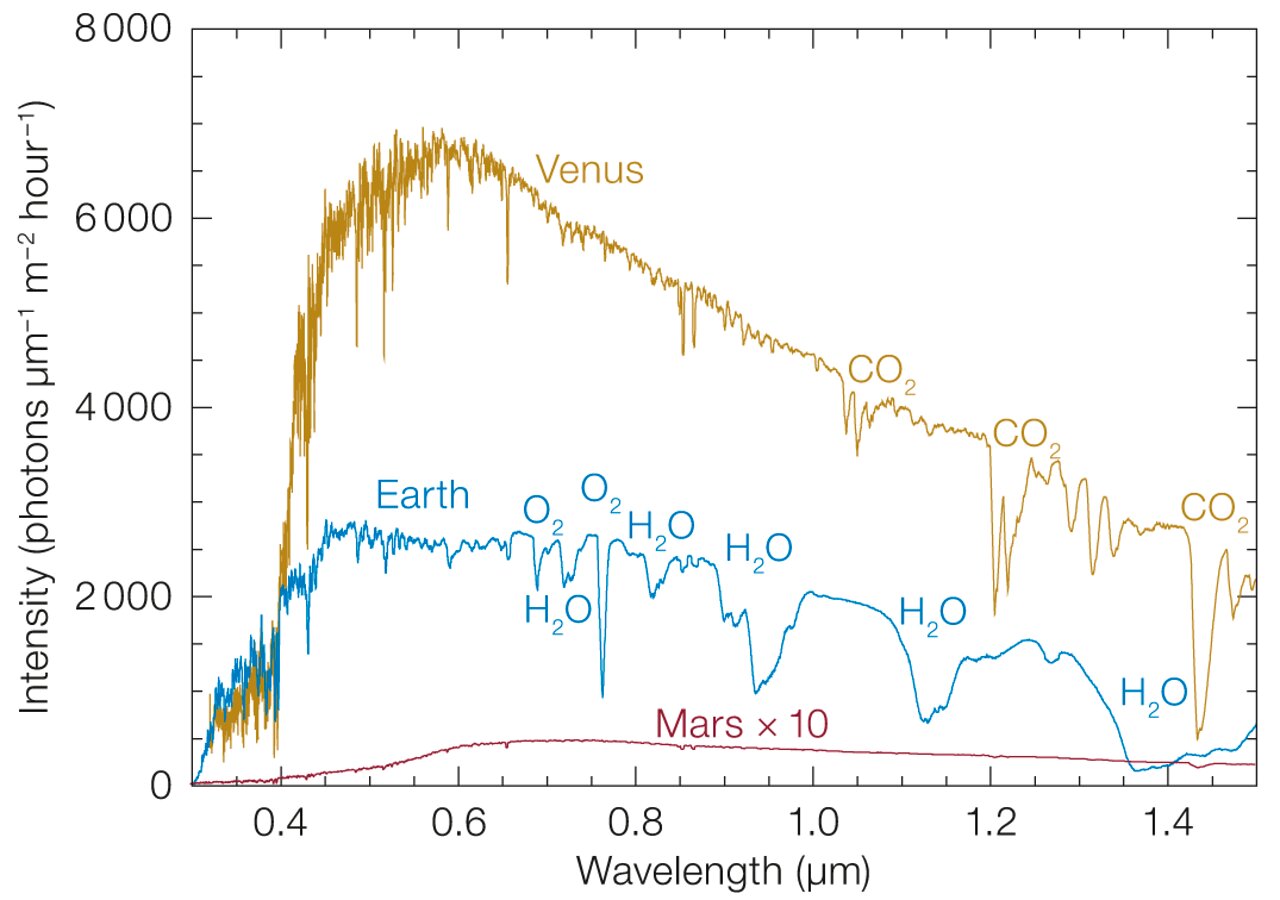}
  \end{center}
  \caption{Spectra of Earth, Venus, and Mars the visible to near
    infrared wavelength range, where the reflected flux of Mars has
    been multiplied by 10 for better visibility
    \cite{kaltenegger2017}.}
  \label{fig:spectra}
\end{figure}

Exoplanet searches to the present date have a limited list of
observable parameters, most of which pertain to the orbital properties
of the planet \cite{pepe2014b}. These observables include orbital
period and eccentricity, and planet mass and radius. The next
significant step forward is expected to lie in the systematic
detection and inventorying of atmospheric compositions via
transmission spectroscopy using the James Webb Space Telescope ({\it
  JWST}) \cite{greene2016,kempton2018}. Though the majority of planets
studied in this manner will likely be gaseous in nature, strategies
have formulated for addressing the challenge of terrestrial planet
atmospheres
\cite{morley2017b,batalha2018b,lincowski2019,lustigyaeger2019a}. Such
methodology will be adapted with improved precision to distinguish
between the spectral signatures of Earth and Venus in an exoplanet
analog context
\cite{ehrenreich2006a,ehrenreich2012a,barstow2016a}. For example,
Figure~\ref{fig:spectra} shows the spectra of Venus, Earth, and Mars
at visible to near infrared (NIR) wavelengths \cite{kaltenegger2017}.
The high optical depth of the Venusian atmosphere results a small
fraction of the upper atmosphere being sampled via transmission
spectroscopy, neglecting the entire troposphere which contains 99\% of
the atmosphere by mass. The CO$_2$ absorption bands expected for Venus
are relatively narrow compared with the broad O$_2$ and H$_2$O
absorption features for Earth's atmosphere, due to the truncation of
absorption features caused the Venusian cloud decks, and the higher
temperature of Earth's atmosphere at high altitudes. Much attention is
directed toward the CO$_2$ absorption bands centered at 2.7 and 4.3
$\mu$m due to their potential detectability with JWST. However, given
the subtle difference in absorption features at those wavelengths for
Venus and Earth analogs, distinguishing between such planets using
transmission spectroscopy may require a broader wavelnegth criteria,
or the identification of Earth-based absorption features such as O$_3$
\cite{barstow2016a,fujii2018}. Since it is expected that many of the
transmission spectroscopy targets from transit surveys will be Venus
analog candidates \cite{kane2014e,lustigyaeger2019b,ostberg2019}, the
disconnect between the available observables, fraction of atmosphere
probed by transmission spectroscopy, and the degeneracy of models that
infer surface conditions, may present a significant barrier to a
complete characterization of terrestrial exoplanets in the near
future.

In the era of exoplanet direct imaging, the reflectance and emission
spectra of terrestrial planets provides an additional means to
characterize their atmospheres
\cite{seager2010,hu2012a,cowan2013a,madhusudhan2019}. Analysis of
disk-integrated reflectance spectra for directly imaged exoplanets
enable access to short wavelength absorption and Rayleigh-scattering
features, similar to those shown in Figure~\ref{fig:spectra}, that can
significantly aid in breaking degeneracy in atmospheric retrieval
models. Thermal emission spectra are measured at longer wavelengths
and reveal the temperature properties of the cloud layers, and
possibly the surface via infrared observing windows. In the context of
Venus, such observational diagnostics are incredibly powerful methods
to probe the deeper atmosphere through observing windows that
penetrate the cloud and haze layers that are largely opaque at optical
wavelengths. NIR observations of the nightside of Venus have
successfully probed beneath the sulfuric acid haze layer to measure
compositions and mixing ratios present in the middle and deep
atmosphere
\cite{bezard1990b,pollack1993b,meadows1996,arney2014}. Remarkably,
these NIR observations also reveal surface emissivity and
topographical features. As described in Section~\ref{venus}, the
temperature, pressure, and compositional profile of the Venusian
atmosphere is critical for a full model evaluation of the complex
climate dynamics and how these relate to the spin state of the
planet. Thus, the combination of transmission, reflectance, and
emission spectra will provide key diagnostics for establishing the
climate state of exoplanets, whose evolution may similarly be related
to the planetary rotation \cite{guzewich2020a}.

\begin{figure*}
  \begin{center}
      \includegraphics[width=0.9\linewidth]{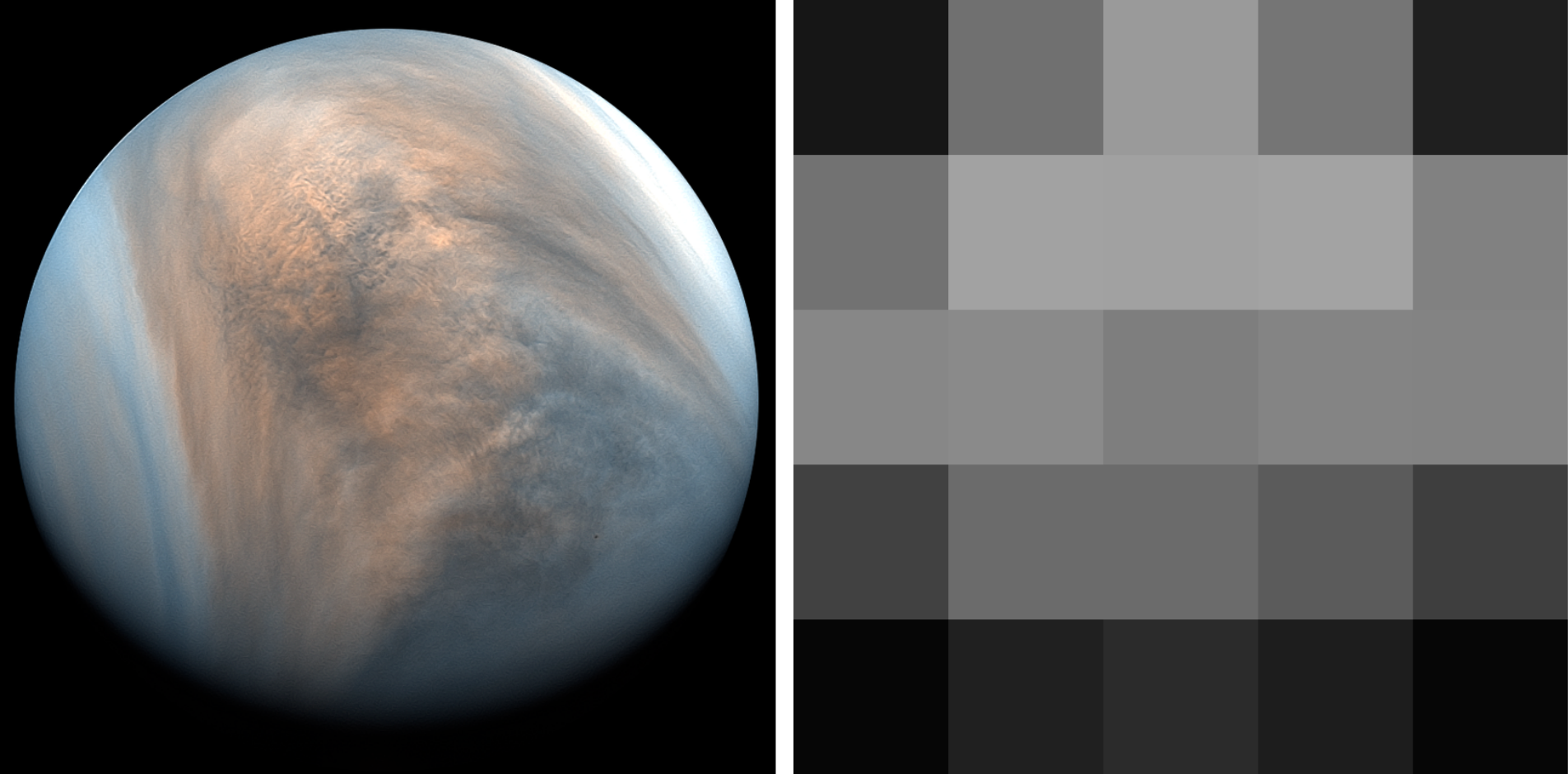}
  \end{center}
  \caption{An image of Venus acquired by the {\it Akatsuki} mission,
    shown in its original high-resolution format (left), and a
    convolved 5$\times$5 pixel format (right) to simulate a direct
    imaging exoplanet observation.}
  \label{fig:imaging}
\end{figure*}

Alternatively, a collaborative approach to exoplanet characterization
between direct and indirect detection techniques may aid in resolving
potential atmospheric model degeneracies. In addition to planetary
mass, radius, and spectroscopy of the upper atmosphere, additional
parameters of rotation, obliquity, atmospheric dynamics, and surface
topography are needed to fully constrain surface conditions. In
particular, Venus atmospheric dynamics serve as a warning with regards
to confusing solid planet with atmospheric rotation, and the
interaction between atmosphere and topography. Shown in the left panel
of Figure~\ref{fig:imaging} is a high-resolution image of Venus
acquired by the {\it Akatsuki} mission. By contrast, the right panel
shows the same data that have been convolved to a 5$\times$5 pixel
image, where a noise model has been included to incorporate systematic
instrumentation effects. Even this image resolution is optimistic for
exoplanet imaging in the short term, and yet the task of extracting
reliable planetary parameters is challenging
\cite{stark2020}. However, the trajectory of future exoplanet mission
planning lies in solving such challenges and directly detecting
terrestrial exoplanets \cite{kopparapu2018}, the vast majority of
which are invisible to the the transit method of detection and
characterization. The Nancy Grace Roman Space Telescope equipped with
a coronagraph will serve as a path-finder and technology demonstration
for achieving exoplanet direct imaging goals
\cite{kasdin2020}. Mission concepts, such as {\it LUVOIR}
\cite{reportluvoir} and the Habitable Exoplanet Observatory (HabEx)
Mission Concept Study \cite{reporthabex}, will be crucial next steps
toward a full terrestrial characterization. The expected limited
signal-to-noise of these data must be convolved with the important
lessons learned from the Venusian atmosphere in order to properly
interpret the dynamics of the plethora of tidally locked planets that
await observation.


\section{Conclusions}
\label{conclusions}

With the dramatic rise in exoplanet discoveries, particularly within
the terrestrial regime, the need to fully characterize the limited
inventory of terrestrial atmospheres within the Solar System is more
important than ever \cite{horner2020b,kane2021d}. Exoplanetary science
is now investing in the task of detecting and interpreting exoplanet
atmospheres, which is a substantial observational and modeling
undertaking. In the near term, many of the exoplanets studied in this
fashion will have been detected via the transit technique. The
consequence of this detection source is a strong bias toward
short-period planets, that will generally have a high chance of both
being tidally locked and pushed into a post runaway greenhouse state,
provided the atmosphere has not been significantly desiccated. As
such, many of the climate simulations conducted for terrestrial
planets have explored the effects of high incident flux and tidal
locking on the possible climate dynamics.

Whilst the exoplanet focus is on Earth-size planets around other
stars, within the Solar System we have an ideal example of an
Earth-sized near tidally locked planet. Venus offers the promise of
in-situ data that we will never have for an exoplanet, and the
possibility of a fully-resolved study of climate dynamics for an
atmosphere that is significantly different from that of the
Earth. Even so, the complexity of the Venusian atmosphere retains many
unanswered questions, such as the composition and chemistry of the
middle and deep atmosphere, and the interaction with the surface. The
opportunity to answer these questions and apply a detailed model of
the Venus climate to other worlds is not only appealing, it is a
necessary step in the quest to characterize the analogs of Earth's
twin around other stars.


\section*{Acknowledgements}

This research has made use of the NASA Exoplanet Archive, which is
operated by the California Institute of Technology under contract with
the National Aeronautics and Space Administration under the Exoplanet
Exploration Program. The results reported herein benefited from
collaborations and/or information exchange within NASA's Nexus for
Exoplanet System Science (NExSS) research coordination network, which
is sponsored by NASA's Science Mission Directorate.


\section*{Data Availability}

Figure~\ref{fig:radii} used data from the NASA Exoplanet Archive,
available here:\\
https://exoplanetarchive.ipac.caltech.edu/ \\
Figure~\ref{fig:rocke3d}
used output data from ROCKE-3D simulations in netCDF format
\cite{kane2018d}. Figure~\ref{fig:imaging} used data from the Akatsuki
Science Data Archive, available here:\\
https://darts.isas.jaxa.jp/planet/project/akatsuki/ \\
The data from figures \ref{fig:radii}, \ref{fig:rocke3d}, and
\ref{fig:imaging} are available
here:\\
http://stephenkane.net/tidalvenus/



\begin{thebibliography}{100}

\bibitem{akeson2013}
R.~L. {Akeson}, X.~{Chen}, D.~{Ciardi}, M.~{Crane}, J.~{Good}, M.~{Harbut},
  E.~{Jackson}, S.~R. {Kane}, A.~C. {Laity}, S.~{Leifer}, M.~{Lynn}, D.~L.
  {McElroy}, M.~{Papin}, P.~{Plavchan}, S.~V. {Ram{\'\i}rez}, R.~{Rey}, K.~{von
  Braun}, M.~{Wittman}, M.~{Abajian}, B.~{Ali}, C.~{Beichman}, A.~{Beekley},
  G.~B. {Berriman}, S.~{Berukoff}, G.~{Bryden}, B.~{Chan}, S.~{Groom},
  C.~{Lau}, A.~N. {Payne}, M.~{Regelson}, M.~{Saucedo}, M.~{Schmitz},
  J.~{Stauffer}, P.~{Wyatt}, and A.~{Zhang}.
\newblock {The NASA Exoplanet Archive: Data and Tools for Exoplanet Research}.
\newblock {\em \pasp}, 125(930):989, Aug 2013.

\bibitem{angelo2017a}
Isabel {Angelo}, Jason~F. {Rowe}, Steve~B. {Howell}, Elisa~V. {Quintana},
  Martin {Still}, Andrew~W. {Mann}, Ben {Burningham}, Thomas {Barclay},
  David~R. {Ciardi}, Daniel {Huber}, and Stephen~R. {Kane}.
\newblock {Kepler-1649b: An Exo-Venus in the Solar Neighborhood}.
\newblock {\em \aj}, 153(4):162, April 2017.

\bibitem{arney2014}
Giada {Arney}, Victoria {Meadows}, David {Crisp}, Sarah~J. {Schmidt}, Jeremy
  {Bailey}, and Tyler {Robinson}.
\newblock {Spatially resolved measurements of H$_{2}$O, HCl, CO, OCS, SO$_{2}$,
  cloud opacity, and acid concentration in the Venus near-infrared spectral
  windows}.
\newblock {\em Journal of Geophysical Research (Planets)}, 119(8):1860--1891,
  August 2014.

\bibitem{auclairdesrotour2020}
P.~{Auclair-Desrotour} and K.~{Heng}.
\newblock {Atmospheric stability and collapse on tidally locked rocky planets}.
\newblock {\em \aap}, 638:A77, June 2020.

\bibitem{auclairdesrotour2017b}
P.~{Auclair-Desrotour}, J.~{Laskar}, S.~{Mathis}, and A.~C.~M. {Correia}.
\newblock {The rotation of planets hosting atmospheric tides: from Venus to
  habitable super-Earths}.
\newblock {\em \aap}, 603:A108, July 2017.

\bibitem{barnes2017b}
Rory {Barnes}.
\newblock {Tidal locking of habitable exoplanets}.
\newblock {\em Celestial Mechanics and Dynamical Astronomy}, 129(4):509--536,
  December 2017.

\bibitem{barnes2013a}
Rory {Barnes}, Kristina {Mullins}, Colin {Goldblatt}, Victoria~S. {Meadows},
  James~F. {Kasting}, and Ren{\'e} {Heller}.
\newblock {Tidal Venuses: Triggering a Climate Catastrophe via Tidal Heating}.
\newblock {\em Astrobiology}, 13(3):225--250, Mar 2013.

\bibitem{barstow2016a}
J.~K. {Barstow}, S.~{Aigrain}, P.~G.~J. {Irwin}, S.~{Kendrew}, and L.~N.
  {Fletcher}.
\newblock {Telling twins apart: exo-Earths and Venuses with transit
  spectroscopy}.
\newblock {\em \mnras}, 458(3):2657--2666, May 2016.

\bibitem{batalha2018b}
Natasha~E. {Batalha}, Nikole~K. {Lewis}, Michael~R. {Line}, Jeff {Valenti}, and
  Kevin {Stevenson}.
\newblock {Strategies for Constraining the Atmospheres of Temperate Terrestrial
  Planets with JWST}.
\newblock {\em \apjl}, 856(2):L34, Apr 2018.

\bibitem{bertaux2016}
Jean-Loup {Bertaux}, I.~V. {Khatuntsev}, A.~{Hauchecorne}, W.~J. {Markiewicz},
  E.~{Marcq}, S.~{Lebonnois}, M.~{Patsaeva}, A.~{Turin}, and A.~{Fedorova}.
\newblock {Influence of Venus topography on the zonal wind and UV albedo at
  cloud top level: The role of stationary gravity waves}.
\newblock {\em Journal of Geophysical Research (Planets)}, 121(6):1087--1101,
  June 2016.

\bibitem{bezard1990b}
B.~{Bezard}, C.~{de Bergh}, D.~{Crisp}, and J.~P. {Maillard}.
\newblock {The deep atmosphere of Venus revealed by high-resolution nightside
  spectra}.
\newblock {\em \nat}, 345(6275):508--511, June 1990.

\bibitem{bills2005e}
Bruce~G. {Bills}.
\newblock {Variations in the rotation rate of Venus due to orbital eccentricity
  modulation of solar tidal torques}.
\newblock {\em Journal of Geophysical Research (Planets)}, 110(E11):E11007, Nov
  2005.

\bibitem{borucki2016}
W.~J. {Borucki}.
\newblock {KEPLER Mission: development and overview}.
\newblock {\em Reports on Progress in Physics}, 79(3):036901, March 2016.

\bibitem{bryson2021}
Steve {Bryson}, Michelle {Kunimoto}, Ravi~K. {Kopparapu}, Jeffrey~L.
  {Coughlin}, William~J. {Borucki}, David {Koch}, Victor~Silva {Aguirre},
  Christopher {Allen}, Geert {Barentsen}, Natalie~M. {Batalha}, Travis
  {Berger}, Alan {Boss}, Lars~A. {Buchhave}, Christopher~J. {Burke}, Douglas~A.
  {Caldwell}, Jennifer~R. {Campbell}, Joseph {Catanzarite}, Hema
  {Chandrasekaran}, William~J. {Chaplin}, Jessie~L. {Christiansen}, J{\o}rgen
  {Christensen-Dalsgaard}, David~R. {Ciardi}, Bruce~D. {Clarke}, William~D.
  {Cochran}, Jessie~L. {Dotson}, Laurance~R. {Doyle}, Eduardo~Seperuelo
  {Duarte}, Edward~W. {Dunham}, Andrea~K. {Dupree}, Michael {Endl}, James~L.
  {Fanson}, Eric~B. {Ford}, Maura {Fujieh}, III {Gautier}, Thomas~N., John~C.
  {Geary}, Ronald~L. {Gilliland}, Forrest~R. {Girouard}, Alan {Gould},
  Michael~R. {Haas}, Christopher~E. {Henze}, Matthew~J. {Holman}, Andrew~W.
  {Howard}, Steve~B. {Howell}, Daniel {Huber}, Roger~C. {Hunter}, Jon~M.
  {Jenkins}, Hans {Kjeldsen}, Jeffery {Kolodziejczak}, Kipp {Larson}, David~W.
  {Latham}, Jie {Li}, Savita {Mathur}, S{\o}ren {Meibom}, Chris {Middour},
  Robert~L. {Morris}, Timothy~D. {Morton}, Fergal {Mullally}, Susan~E.
  {Mullally}, David {Pletcher}, Andrej {Prsa}, Samuel~N. {Quinn}, Elisa~V.
  {Quintana}, Darin {Ragozzine}, Solange~V. {Ramirez}, Dwight~T. {Sanderfer},
  Dimitar {Sasselov}, Shawn~E. {Seader}, Megan {Shabram}, Avi {Shporer},
  Jeffrey~C. {Smith}, Jason~H. {Steffen}, Martin {Still}, Guillermo {Torres},
  John {Troeltzsch}, Joseph~D. {Twicken}, Akm~Kamal {Uddin}, Jeffrey~E. {Van
  Cleve}, Janice {Voss}, Lauren~M. {Weiss}, William~F. {Welsh}, Bill {Wohler},
  and Khadeejah~A. {Zamudio}.
\newblock {The Occurrence of Rocky Habitable-zone Planets around Solar-like
  Stars from Kepler Data}.
\newblock {\em \aj}, 161(1):36, January 2021.

\bibitem{butler2006}
R.~P. {Butler}, J.~T. {Wright}, G.~W. {Marcy}, D.~A. {Fischer}, S.~S. {Vogt},
  C.~G. {Tinney}, H.~R.~A. {Jones}, B.~D. {Carter}, J.~A. {Johnson},
  C.~{McCarthy}, and A.~J. {Penny}.
\newblock {Catalog of Nearby Exoplanets}.
\newblock {\em \apj}, 646:505--522, July 2006.

\bibitem{campbell2019}
Bruce~A. {Campbell}, Donald~B. {Campbell}, Lynn~M. {Carter}, John~F.
  {Chandler}, Jon~D. {Giorgini}, Jean-Luc {Margot}, Gareth~A. {Morgan},
  Michael~C. {Nolan}, Phillip~J. {Perillat}, and Jennifer~L. {Whitten}.
\newblock {The mean rotation rate of Venus from 29 years of Earth-based radar
  observations}.
\newblock {\em \icarus}, 332:19--23, November 2019.

\bibitem{carone2016}
L.~{Carone}, R.~{Keppens}, and L.~{Decin}.
\newblock {Connecting the dots - III. Nightside cooling and surface friction
  affect climates of tidally locked terrestrial planets}.
\newblock {\em \mnras}, 461(2):1981--2002, September 2016.

\bibitem{christensen2006}
U.~R. {Christensen} and J.~{Aubert}.
\newblock {Scaling properties of convection-driven dynamos in rotating
  spherical shells and application to planetary magnetic fields}.
\newblock {\em Geophysical Journal International}, 166(1):97--114, July 2006.

\bibitem{clanton2014b}
Christian {Clanton} and B.~Scott {Gaudi}.
\newblock {Synthesizing Exoplanet Demographics from Radial Velocity and
  Microlensing Surveys. II. The Frequency of Planets Orbiting M Dwarfs}.
\newblock {\em \apj}, 791(2):91, August 2014.

\bibitem{correia2001}
Alexandre C.~M. {Correia} and Jacques {Laskar}.
\newblock {The four final rotation states of Venus}.
\newblock {\em \nat}, 411(6839):767--770, June 2001.

\bibitem{cowan2013a}
Nicolas~B. {Cowan} and Talia~E. {Strait}.
\newblock {Determining Reflectance Spectra of Surfaces and Clouds on
  Exoplanets}.
\newblock {\em \apjl}, 765(1):L17, March 2013.

\bibitem{crisp1986a}
D.~{Crisp}.
\newblock {Radiative forcing of the Venus mesosphere I. Solar fluxes and
  heating rates}.
\newblock {\em \icarus}, 67(3):484--514, September 1986.

\bibitem{delgenio2019a}
Anthony~D. {Del Genio}, Michael~J. {Way}, David~S. {Amundsen}, Igor {Aleinov},
  Maxwell {Kelley}, Nancy~Y. {Kiang}, and Thomas~L. {Clune}.
\newblock {Habitable Climate Scenarios for Proxima Centauri b with a Dynamic
  Ocean}.
\newblock {\em Astrobiology}, 19(1):99--125, January 2019.

\bibitem{ding2020a}
Feng {Ding} and Robin~D. {Wordsworth}.
\newblock {Stabilization of Dayside Surface Liquid Water via Tropopause Cold
  Trapping on Arid Slowly Rotating Tidally Locked Planets}.
\newblock {\em \apjl}, 891(1):L18, March 2020.

\bibitem{dressing2013}
Courtney~D. {Dressing} and David {Charbonneau}.
\newblock {The Occurrence Rate of Small Planets around Small Stars}.
\newblock {\em \apj}, 767(1):95, Apr 2013.

\bibitem{driscoll2015}
P.~E. {Driscoll} and R.~{Barnes}.
\newblock {Tidal Heating of Earth-like Exoplanets around M Stars: Thermal,
  Magnetic, and Orbital Evolutions}.
\newblock {\em Astrobiology}, 15(9):739--760, Sep 2015.

\bibitem{dumoulin2017}
C.~{Dumoulin}, G.~{Tobie}, O.~{Verhoeven}, P.~{Rosenblatt}, and N.~{Rambaux}.
\newblock {Tidal constraints on the interior of Venus}.
\newblock {\em Journal of Geophysical Research (Planets)}, 122(6):1338--1352,
  Jun 2017.

\bibitem{ehrenreich2006a}
D.~{Ehrenreich}, G.~{Tinetti}, A.~{Lecavelier Des Etangs}, A.~{Vidal-Madjar},
  and F.~{Selsis}.
\newblock {The transmission spectrum of Earth-size transiting planets}.
\newblock {\em \aap}, 448(1):379--393, March 2006.

\bibitem{ehrenreich2012a}
D.~{Ehrenreich}, A.~{Vidal-Madjar}, T.~{Widemann}, G.~{Gronoff}, P.~{Tanga},
  M.~{Barth{\'e}lemy}, J.~{Lilensten}, A.~{Lecavelier Des Etangs}, and
  L.~{Arnold}.
\newblock {Transmission spectrum of Venus as a transiting exoplanet}.
\newblock {\em \aap}, 537:L2, Jan 2012.

\bibitem{fauchez2021}
Thomas~J. {Fauchez}, Martin {Turbet}, Denis~E. {Sergeev}, Nathan~J. {Mayne},
  Aymeric {Spiga}, Linda {Sohl}, Prabal {Saxena}, Russell {Deitrick}, Gabriella
  {Gilli}, Shawn~D. {Domagal-Goldman}, Fran{\c{c}}cois {Forget}, Richard
  {Consentino}, Rory {Barnes}, Jacob {Haqq-Misra}, M.~J. {Way}, Eric~T. {Wolf},
  Stephanie {Olson}, Jaime~S. {Crouse}, Estelle {Janin}, Emeline {Bolmont},
  J{\'e}r{\'e}my {Leconte}, Guillaume {Chaverot}, Yassin {Jaziri}, Kostantinos
  {Tsigaridis}, Jun {Yang}, Daria {Pidhorodetska}, Ravi~K. {Kopparapu}, Howard
  {Chen}, Ian~A. {Boutle}, Maxence {Lefevre}, Benjamin {Charnay}, Andy
  {Burnett}, John {Cabra}, and Najja {Bouldin}.
\newblock {TRAPPIST Habitable Atmosphere Intercomparison (THAI) Workshop
  Report}.
\newblock {\em \psj}, 2(3):106, June 2021.

\bibitem{ford2014}
Eric~B. {Ford}.
\newblock {Architectures of planetary systems and implications for their
  formation}.
\newblock {\em Proceedings of the National Academy of Science},
  111(35):12616--12621, Sep 2014.

\bibitem{forget2014}
F.~{Forget} and J.~{Leconte}.
\newblock {Possible climates on terrestrial exoplanets}.
\newblock {\em Philosophical Transactions of the Royal Society of London Series
  A}, 372(2014):20130084--20130084, March 2014.

\bibitem{fujii2018}
Yuka {Fujii}, Daniel {Angerhausen}, Russell {Deitrick}, Shawn
  {Domagal-Goldman}, John~Lee {Grenfell}, Yasunori {Hori}, Stephen~R. {Kane},
  Enric {Pall{\'e}}, Heike {Rauer}, Nicholas {Siegler}, Karl {Stapelfeldt}, and
  Kevin~B. {Stevenson}.
\newblock {Exoplanet Biosignatures: Observational Prospects}.
\newblock {\em Astrobiology}, 18(6):739--778, June 2018.

\bibitem{fukuhara2017a}
Tetsuya {Fukuhara}, Masahiko {Futaguchi}, George~L. {Hashimoto}, Takeshi
  {Horinouchi}, Takeshi {Imamura}, Naomoto {Iwagaimi}, Toru {Kouyama}, Shin-Ya
  {Murakami}, Masato {Nakamura}, Kazunori {Ogohara}, Mitsuteru {Sato}, Takao~M.
  {Sato}, Makoto {Suzuki}, Makoto {Taguchi}, Seiko {Takagi}, Munetaka {Ueno},
  Shigeto {Watanabe}, Manabu {Yamada}, and Atsushi {Yamazaki}.
\newblock {Large stationary gravity wave in the atmosphere of Venus}.
\newblock {\em Nature Geoscience}, 10(2):85--88, Jan 2017.

\bibitem{fulton2017}
Benjamin~J. {Fulton}, Erik~A. {Petigura}, Andrew~W. {Howard}, Howard
  {Isaacson}, Geoffrey~W. {Marcy}, Phillip~A. {Cargile}, Leslie {Hebb},
  Lauren~M. {Weiss}, John~Asher {Johnson}, Timothy~D. {Morton}, Evan
  {Sinukoff}, Ian J.~M. {Crossfield}, and Lea~A. {Hirsch}.
\newblock {The California-Kepler Survey. III. A Gap in the Radius Distribution
  of Small Planets}.
\newblock {\em \aj}, 154(3):109, Sep 2017.

\bibitem{funk2010}
B.~{Funk}, G.~{Wuchterl}, R.~{Schwarz}, E.~{Pilat-Lohinger}, and S.~{Eggl}.
\newblock {The stability of ultra-compact planetary systems}.
\newblock {\em \aap}, 516:A82, June 2010.

\bibitem{reporthabex}
B.~Scott {Gaudi}, Sara {Seager}, Bertrand {Mennesson}, Alina {Kiessling}, Keith
  {Warfield}, Kerri {Cahoy}, John~T. {Clarke}, Shawn {Domagal-Goldman}, Lee
  {Feinberg}, Olivier {Guyon}, Jeremy {Kasdin}, Dimitri {Mawet}, Peter
  {Plavchan}, Tyler {Robinson}, Leslie {Rogers}, Paul {Scowen}, Rachel
  {Somerville}, Karl {Stapelfeldt}, Christopher {Stark}, Daniel {Stern},
  Margaret {Turnbull}, Rashied {Amini}, Gary {Kuan}, Stefan {Martin}, Rhonda
  {Morgan}, David {Redding}, H.~Philip {Stahl}, Ryan {Webb}, Oscar
  {Alvarez-Salazar}, William~L. {Arnold}, Manan {Arya}, Bala {Balasubramanian},
  Mike {Baysinger}, Ray {Bell}, Chris {Below}, Jonathan {Benson}, Lindsey
  {Blais}, Jeff {Booth}, Robert {Bourgeois}, Case {Bradford}, Alden {Brewer},
  Thomas {Brooks}, Eric {Cady}, Mary {Caldwell}, Rob {Calvet}, Steven {Carr},
  Derek {Chan}, Velibor {Cormarkovic}, Keith {Coste}, Charlie {Cox}, Rolf
  {Danner}, Jacqueline {Davis}, Larry {Dewell}, Lisa {Dorsett}, Daniel {Dunn},
  Matthew {East}, Michael {Effinger}, Ron {Eng}, Greg {Freebury}, Jay {Garcia},
  Jonathan {Gaskin}, Suzan {Greene}, John {Hennessy}, Evan {Hilgemann}, Brad
  {Hood}, Wolfgang {Holota}, Scott {Howe}, Pei {Huang}, Tony {Hull}, Ron
  {Hunt}, Kevin {Hurd}, Sandra {Johnson}, Andrew {Kissil}, Brent {Knight},
  Daniel {Kolenz}, Oliver {Kraus}, John {Krist}, Mary {Li}, Doug {Lisman},
  Milan {Mandic}, John {Mann}, Luis {Marchen}, Colleen {Marrese-Reading},
  Jonathan {McCready}, Jim {McGown}, Jessica {Missun}, Andrew {Miyaguchi},
  Bradley {Moore}, Bijan {Nemati}, Shouleh {Nikzad}, Joel {Nissen}, Megan
  {Novicki}, Todd {Perrine}, Claudia {Pineda}, Otto {Polanco}, Dustin {Putnam},
  Atif {Qureshi}, Michael {Richards}, A.~J. {Eldorado Riggs}, Michael
  {Rodgers}, Mike {Rud}, Navtej {Saini}, Dan {Scalisi}, Dan {Scharf}, Kevin
  {Schulz}, Gene {Serabyn}, Norbert {Sigrist}, Glory {Sikkia}, Andrew
  {Singleton}, Stuart {Shaklan}, Scott {Smith}, Bart {Southerd}, Mark {Stahl},
  John {Steeves}, Brian {Sturges}, Chris {Sullivan}, Hao {Tang}, Neil {Taras},
  Jonathan {Tesch}, Melissa {Therrell}, Howard {Tseng}, Marty {Valente}, David
  {Van Buren}, Juan {Villalvazo}, Steve {Warwick}, David {Webb}, Thomas
  {Westerhoff}, Rush {Wofford}, Gordon {Wu}, Jahning {Woo}, Milana {Wood}, John
  {Ziemer}, Giada {Arney}, Jay {Anderson}, Jes{\'u}s {Ma{\'\i}z-Apell{\'a}niz},
  James {Bartlett}, Ruslan {Belikov}, Eduardo {Bendek}, Brad {Cenko}, Ewan
  {Douglas}, Shannon {Dulz}, Chris {Evans}, Virginie {Faramaz}, Y.~Katherina
  {Feng}, Harry {Ferguson}, Kate {Follette}, Saavik {Ford}, Miriam
  {Garc{\'\i}a}, Marla {Geha}, Dawn {Gelino}, Ylva {G{\"o}tberg}, Sergi
  {Hildebrand t}, Renyu {Hu}, Knud {Jahnke}, Grant {Kennedy}, Laura
  {Kreidberg}, Andrea {Isella}, Eric {Lopez}, Franck {Marchis}, Lucas {Macri},
  Mark {Marley}, William {Matzko}, Johan {Mazoyer}, Stephan {McCandliss},
  Tiffany {Meshkat}, Christoph {Mordasini}, Patrick {Morris}, Eric {Nielsen},
  Patrick {Newman}, Erik {Petigura}, Marc {Postman}, Amy {Reines}, Aki
  {Roberge}, Ian {Roederer}, Garreth {Ruane}, Edouard {Schwieterman}, Dan
  {Sirbu}, Christopher {Spalding}, Harry {Teplitz}, Jason {Tumlinson}, Neal
  {Turner}, Jessica {Werk}, Aida {Wofford}, Mark {Wyatt}, Amber {Young}, and
  Rob {Zellem}.
\newblock {The Habitable Exoplanet Observatory (HabEx) Mission Concept Study
  Final Report}.
\newblock {\em arXiv e-prints}, page arXiv:2001.06683, January 2020.

\bibitem{gladman1996b}
Brett {Gladman}, D.~Dane {Quinn}, Philip {Nicholson}, and Richard {Rand}.
\newblock {Synchronous Locking of Tidally Evolving Satellites}.
\newblock {\em \icarus}, 122(1):166--192, July 1996.

\bibitem{greene2016}
Thomas~P. {Greene}, Michael~R. {Line}, Cezar {Montero}, Jonathan~J. {Fortney},
  Jacob {Lustig-Yaeger}, and Kyle {Luther}.
\newblock {Characterizing Transiting Exoplanet Atmospheres with JWST}.
\newblock {\em \apj}, 817(1):17, January 2016.

\bibitem{gunell2018a}
Herbert {Gunell}, Romain {Maggiolo}, Hans {Nilsson}, Gabriella {Stenberg
  Wieser}, Rikard {Slapak}, Jesper {Lindkvist}, Maria {Hamrin}, and Johan {De
  Keyser}.
\newblock {Why an intrinsic magnetic field does not protect a planet against
  atmospheric escape}.
\newblock {\em \aap}, 614:L3, June 2018.

\bibitem{guzewich2020a}
Scott~D. {Guzewich}, Jacob {Lustig-Yaeger}, Christopher~Evan {Davis},
  Ravi~Kumar {Kopparapu}, Michael~J. {Way}, and Victoria~S. {Meadows}.
\newblock {The Impact of Planetary Rotation Rate on the Reflectance and Thermal
  Emission Spectrum of Terrestrial Exoplanets around Sunlike Stars}.
\newblock {\em \apj}, 893(2):140, April 2020.

\bibitem{hamano2013}
Keiko {Hamano}, Yutaka {Abe}, and Hidenori {Genda}.
\newblock {Emergence of two types of terrestrial planet on solidification of
  magma ocean}.
\newblock {\em \nat}, 497(7451):607--610, May 2013.

\bibitem{hammond2021}
Mark {Hammond} and Neil~T. {Lewis}.
\newblock {The rotational and divergent components of atmospheric circulation
  on tidally locked planets}.
\newblock {\em Proceedings of the National Academy of Science},
  118(13):2022705118, March 2021.

\bibitem{horinouchi2020}
Takeshi {Horinouchi}, Yoshi-Yuki {Hayashi}, Shigeto {Watanabe}, Manabu
  {Yamada}, Atsushi {Yamazaki}, Toru {Kouyama}, Makoto {Taguchi}, Tetsuya
  {Fukuhara}, Masahiro {Takagi}, Kazunori {Ogohara}, Shin-ya {Murakami}, Javier
  {Peralta}, Sanjay~S. {Limaye}, Takeshi {Imamura}, Masato {Nakamura}, Takao~M.
  {Sato}, and Takehiko {Satoh}.
\newblock {How waves and turbulence maintain the super-rotation of
  Venus{\textquoteright} atmosphere}.
\newblock {\em Science}, 368(6489):405--409, April 2020.

\bibitem{horner2020b}
J.~{Horner}, S.~R. {Kane}, J.~P. {Marshall}, P.~A. {Dalba}, T.~R. {Holt},
  J.~{Wood}, H.~E. {Maynard-Casely}, R.~{Wittenmyer}, P.~S. {Lykawka},
  M.~{Hill}, R.~{Salmeron}, J.~{Bailey}, T.~{L{\"o}hne}, M.~{Agnew}, B.~D.
  {Carter}, and C.~C.~E. {Tylor}.
\newblock {Solar System Physics for Exoplanet Research}.
\newblock {\em \pasp}, 132(1016):102001, October 2020.

\bibitem{hu2012a}
Renyu {Hu}, Bethany~L. {Ehlmann}, and Sara {Seager}.
\newblock {Theoretical Spectra of Terrestrial Exoplanet Surfaces}.
\newblock {\em \apj}, 752(1):7, June 2012.

\bibitem{imamura2020}
Takeshi {Imamura}, Jonathan {Mitchell}, Sebastien {Lebonnois}, Yohai {Kaspi},
  Adam~P. {Showman}, and Oleg {Korablev}.
\newblock {Superrotation in Planetary Atmospheres}.
\newblock {\em \ssr}, 216(5):87, July 2020.

\bibitem{ingersoll1978b}
A.~P. {Ingersoll} and A.~R. {Dobrovolskis}.
\newblock {Venus' rotation and atmospheric tides}.
\newblock {\em \nat}, 275(5675):37--38, Sep 1978.

\bibitem{kaltenegger2017}
Lisa {Kaltenegger}.
\newblock {How to Characterize Habitable Worlds and Signs of Life}.
\newblock {\em \araa}, 55(1):433--485, August 2017.

\bibitem{kane2018d}
S.~R. {Kane}, A.~Y. {Ceja}, M.~J. {Way}, and E.~V. {Quintana}.
\newblock {Climate Modeling of a Potential ExoVenus}.
\newblock {\em \apj}, 869:46, December 2018.

\bibitem{kane2012a}
S.~R. {Kane} and D.~M. {Gelino}.
\newblock {The Habitable Zone Gallery}.
\newblock {\em \pasp}, 124:323, April 2012.

\bibitem{kane2016c}
S.~R. {Kane}, M.~L. {Hill}, J.~F. {Kasting}, R.~K. {Kopparapu}, E.~V.
  {Quintana}, T.~{Barclay}, N.~M. {Batalha}, W.~J. {Borucki}, D.~R. {Ciardi},
  N.~{Haghighipour}, N.~R. {Hinkel}, L.~{Kaltenegger}, F.~{Selsis}, and
  G.~{Torres}.
\newblock {A Catalog of Kepler Habitable Zone Exoplanet Candidates}.
\newblock {\em \apj}, 830:1, October 2016.

\bibitem{kane2013e}
S.~R. {Kane}, N.~R. {Hinkel}, and S.~N. {Raymond}.
\newblock {Solar System Moons as Analogs for Compact Exoplanetary Systems}.
\newblock {\em \aj}, 146:122, November 2013.

\bibitem{kane2014e}
S.~R. {Kane}, R.~K. {Kopparapu}, and S.~D. {Domagal-Goldman}.
\newblock {On the Frequency of Potential Venus Analogs from Kepler Data}.
\newblock {\em \apjl}, 794:L5, October 2014.

\bibitem{kane2019d}
Stephen~R. {Kane}, Giada {Arney}, David {Crisp}, Shawn {Domagal-Goldman},
  Lori~S. {Glaze}, Colin {Goldblatt}, David {Grinspoon}, James~W. {Head},
  Adrian {Lenardic}, Cayman {Unterborn}, Michael~J. {Way}, and Kevin~J.
  {Zahnle}.
\newblock {Venus as a Laboratory for Exoplanetary Science}.
\newblock {\em Journal of Geophysical Research (Planets)}, 124(8):2015--2028,
  Aug 2019.

\bibitem{kane2021d}
Stephen~R. {Kane}, Giada~N. {Arney}, Paul~K. {Byrne}, Paul~A. {Dalba},
  Steven~J. {Desch}, Jonti {Horner}, Noam~R. {Izenberg}, Kathleen~E. {Mandt},
  Victoria~S. {Meadows}, and Lynnae~C. {Quick}.
\newblock {The Fundamental Connections between the Solar System and
  Exoplanetary Science}.
\newblock {\em Journal of Geophysical Research (Planets)}, 126(2):e06643,
  February 2021.

\bibitem{kane2020e}
Stephen~R. {Kane}, Pam {Vervoort}, Jonathan {Horner}, and Francisco~J.
  {Pozuelos}.
\newblock {Could the Migration of Jupiter Have Accelerated the Atmospheric
  Evolution of Venus?}
\newblock {\em The Planetary Science Journal}, 1(2):42, September 2020.

\bibitem{kasdin2020}
N.~Jeremy {Kasdin}, Vanessa~P. {Bailey}, Bertrand {Mennesson}, Robert~T.
  {Zellem}, Marie {Ygouf}, Jason {Rhodes}, Thomas {Luchik}, Feng {Zhao},
  A.~J.~Eldorado {Riggs}, Byoung-Joon {Seo}, John {Krist}, Brian {Kern}, Hong
  {Tang}, Bijan {Nemati}, Tyler~D. {Groff}, Neil {Zimmerman}, Bruce
  {Macintosh}, Margaret {Turnbull}, John {Debes}, Ewan~S. {Douglas}, and
  Roxana~E. {Lupu}.
\newblock {The Nancy Grace Roman Space Telescope Coronagraph Instrument (CGI)
  technology demonstration}.
\newblock In {\em Society of Photo-Optical Instrumentation Engineers (SPIE)
  Conference Series}, volume 11443 of {\em Society of Photo-Optical
  Instrumentation Engineers (SPIE) Conference Series}, page 114431U, December
  2020.

\bibitem{kaspi2020}
Yohai {Kaspi}, Eli {Galanti}, Adam~P. {Showman}, David~J. {Stevenson}, Tristan
  {Guillot}, Luciano {Iess}, and Scott~J. {Bolton}.
\newblock {Comparison of the Deep Atmospheric Dynamics of Jupiter and Saturn in
  Light of the Juno and Cassini Gravity Measurements}.
\newblock {\em \ssr}, 216(5):84, June 2020.

\bibitem{kasting1988c}
J.~F. {Kasting}.
\newblock {Runaway and moist greenhouse atmospheres and the evolution of Earth
  and Venus}.
\newblock {\em \icarus}, 74(3):472--494, Jun 1988.

\bibitem{kasting1993a}
James~F. {Kasting}, Daniel~P. {Whitmire}, and Ray~T. {Reynolds}.
\newblock {Habitable Zones around Main Sequence Stars}.
\newblock {\em \icarus}, 101(1):108--128, Jan 1993.

\bibitem{kempton2018}
Eliza M.~R. {Kempton}, Jacob~L. {Bean}, Dana~R. {Louie}, Drake {Deming}, Daniel
  D.~B. {Koll}, Megan {Mansfield}, Jessie~L. {Christiansen}, Mercedes
  {L{\'o}pez-Morales}, Mark~R. {Swain}, Robert~T. {Zellem}, Sarah {Ballard},
  Thomas {Barclay}, Joanna~K. {Barstow}, Natasha~E. {Batalha}, Thomas~G.
  {Beatty}, Zach {Berta-Thompson}, Jayne {Birkby}, Lars~A. {Buchhave}, David
  {Charbonneau}, Nicolas~B. {Cowan}, Ian {Crossfield}, Miguel {de Val-Borro},
  Ren{\'e} {Doyon}, Diana {Dragomir}, Eric {Gaidos}, Kevin {Heng}, Renyu {Hu},
  Stephen~R. {Kane}, Laura {Kreidberg}, Matthias {Mallonn}, Caroline~V.
  {Morley}, Norio {Narita}, Valerio {Nascimbeni}, Enric {Pall{\'e}}, Elisa~V.
  {Quintana}, Emily {Rauscher}, Sara {Seager}, Evgenya~L. {Shkolnik}, David~K.
  {Sing}, Alessandro {Sozzetti}, Keivan~G. {Stassun}, Jeff~A. {Valenti}, and
  Carolina {von Essen}.
\newblock {A Framework for Prioritizing the TESS Planetary Candidates Most
  Amenable to Atmospheric Characterization}.
\newblock {\em \pasp}, 130(993):114401, Nov 2018.

\bibitem{kite2011c}
Edwin~S. {Kite}, Eric {Gaidos}, and Michael {Manga}.
\newblock {Climate Instability on Tidally Locked Exoplanets}.
\newblock {\em \apj}, 743(1):41, December 2011.

\bibitem{koll2016}
Daniel D.~B. {Koll} and Dorian~S. {Abbot}.
\newblock {Temperature Structure and Atmospheric Circulation of Dry Tidally
  Locked Rocky Exoplanets}.
\newblock {\em \apj}, 825(2):99, July 2016.

\bibitem{konopliv1996}
A.~S. {Konopliv} and C.~F. {Yoder}.
\newblock {Venusian k $_{2}$ tidal Love number from Magellan and PVO tracking
  data}.
\newblock {\em \grl}, 23(14):1857--1860, Jan 1996.

\bibitem{kopparapu2018}
R.~K. {Kopparapu}, E.~{H{\'e}brard}, R.~{Belikov}, N.~M. {Batalha}, G.~D.
  {Mulders}, C.~{Stark}, D.~{Teal}, S.~{Domagal-Goldman}, and A.~{Mandell}.
\newblock {Exoplanet Classification and Yield Estimates for Direct Imaging
  Missions}.
\newblock {\em \apj}, 856:122, April 2018.

\bibitem{kopparapu2013a}
Ravi~Kumar {Kopparapu}, Ramses {Ramirez}, James~F. {Kasting}, Vincent {Eymet},
  Tyler~D. {Robinson}, Suvrath {Mahadevan}, Ryan~C. {Terrien}, Shawn
  {Domagal-Goldman}, Victoria {Meadows}, and Rohit {Deshpande}.
\newblock {Habitable Zones around Main-sequence Stars: New Estimates}.
\newblock {\em \apj}, 765(2):131, Mar 2013.

\bibitem{kopparapu2014}
Ravi~Kumar {Kopparapu}, Ramses~M. {Ramirez}, James {SchottelKotte}, James~F.
  {Kasting}, Shawn {Domagal-Goldman}, and Vincent {Eymet}.
\newblock {Habitable Zones around Main-sequence Stars: Dependence on Planetary
  Mass}.
\newblock {\em \apj}, 787(2):L29, Jun 2014.

\bibitem{kouyama2017}
T.~{Kouyama}, T.~{Imamura}, M.~{Taguchi}, T.~{Fukuhara}, T.~M. {Sato},
  A.~{Yamazaki}, M.~{Futaguchi}, S.~{Murakami}, G.~L. {Hashimoto}, M.~{Ueno},
  N.~{Iwagami}, S.~{Takagi}, M.~{Takagi}, K.~{Ogohara}, H.~{Kashimura},
  T.~{Horinouchi}, N.~{Sato}, M.~{Yamada}, Y.~{Yamamoto}, S.~{Ohtsuki},
  K.~{Sugiyama}, H.~{Ando}, M.~{Takamura}, T.~{Yamada}, T.~{Satoh}, and
  M.~{Nakamura}.
\newblock {Topographical and Local Time Dependence of Large Stationary Gravity
  Waves Observed at the Cloud Top of Venus}.
\newblock {\em \grl}, 44(24):12,098--12,105, December 2017.

\bibitem{lebonnois2010}
S{\'e}bastien {Lebonnois}, Fr{\'e}d{\'e}ric {Hourdin}, Vincent {Eymet}, Audrey
  {Crespin}, Richard {Fournier}, and Fran{\c{c}}ois {Forget}.
\newblock {Superrotation of Venus' atmosphere analyzed with a full general
  circulation model}.
\newblock {\em Journal of Geophysical Research (Planets)}, 115(E6):E06006, June
  2010.

\bibitem{lebonnois2020a}
S{\'e}bastien {Lebonnois}, Gerald {Schubert}, Tibor {Kremic}, Leah~M. {Nakley},
  Kyle~G. {Phillips}, Josette {Bellan}, and Daniel {Cordier}.
\newblock {An experimental study of the mixing of CO$_{2}$ and N$_{2}$ under
  conditions found at the surface of Venus}.
\newblock {\em \icarus}, 338:113550, March 2020.

\bibitem{leconte2015a}
J{\'e}r{\'e}my {Leconte}, Hanbo {Wu}, Kristen {Menou}, and Norman {Murray}.
\newblock {Asynchronous rotation of Earth-mass planets in the habitable zone of
  lower-mass stars}.
\newblock {\em Science}, 347(6222):632--635, February 2015.

\bibitem{lee2007c}
C.~{Lee}, S.~R. {Lewis}, and P.~L. {Read}.
\newblock {Superrotation in a Venus general circulation model}.
\newblock {\em Journal of Geophysical Research (Planets)}, 112(E4):E04S11,
  April 2007.

\bibitem{lewis2018}
Neil~T. {Lewis}, F.~Hugo {Lambert}, Ian~A. {Boutle}, Nathan~J. {Mayne}, James
  {Manners}, and David~M. {Acreman}.
\newblock {The Influence of a Substellar Continent on the Climate of a Tidally
  Locked Exoplanet}.
\newblock {\em \apj}, 854(2):171, February 2018.

\bibitem{lincowski2019}
Andrew~P. {Lincowski}, Jacob {Lustig-Yaeger}, and Victoria~S. {Meadows}.
\newblock {Observing Isotopologue Bands in Terrestrial Exoplanet Atmospheres
  with the James Webb Space Telescope: Implications for Identifying Past
  Atmospheric and Ocean Loss}.
\newblock {\em \aj}, 158(1):26, July 2019.

\bibitem{lopez2013}
Eric~D. {Lopez} and Jonathan~J. {Fortney}.
\newblock {The Role of Core Mass in Controlling Evaporation: The Kepler Radius
  Distribution and the Kepler-36 Density Dichotomy}.
\newblock {\em \apj}, 776(1):2, October 2013.

\bibitem{lustigyaeger2019b}
Jacob {Lustig-Yaeger}, Victoria~S. {Meadows}, and Andrew~P. {Lincowski}.
\newblock {A Mirage of the Cosmic Shoreline: Venus-like Clouds as a Statistical
  False Positive for Exoplanet Atmospheric Erosion}.
\newblock {\em \apjl}, 887(1):L11, December 2019.

\bibitem{lustigyaeger2019a}
Jacob {Lustig-Yaeger}, Victoria~S. {Meadows}, and Andrew~P. {Lincowski}.
\newblock {The Detectability and Characterization of the TRAPPIST-1 Exoplanet
  Atmospheres with JWST}.
\newblock {\em \aj}, 158(1):27, July 2019.

\bibitem{madhusudhan2019}
Nikku {Madhusudhan}.
\newblock {Exoplanetary Atmospheres: Key Insights, Challenges, and Prospects}.
\newblock {\em \araa}, 57:617--663, August 2019.

\bibitem{meadows1996}
V.~S. {Meadows} and D.~{Crisp}.
\newblock {Ground-based near-infrared observations of the Venus nightside: The
  thermal structure and water abundance near the surface}.
\newblock {\em \jgr}, 101(E2):4595--4622, January 1996.

\bibitem{morley2017b}
Caroline~V. {Morley}, Laura {Kreidberg}, Zafar {Rustamkulov}, Tyler {Robinson},
  and Jonathan~J. {Fortney}.
\newblock {Observing the Atmospheres of Known Temperate Earth-sized Planets
  with JWST}.
\newblock {\em \apj}, 850(2):121, December 2017.

\bibitem{navarro2018}
T.~{Navarro}, G.~{Schubert}, and S.~{Lebonnois}.
\newblock {Atmospheric mountain wave generation on Venus and its influence on
  the solid planet's rotation rate}.
\newblock {\em Nature Geoscience}, 11(7):487--491, June 2018.

\bibitem{ostberg2019}
Colby {Ostberg} and Stephen~R. {Kane}.
\newblock {Predicting the Yield of Potential Venus Analogs from TESS and Their
  Potential for Atmospheric Characterization}.
\newblock {\em \aj}, 158(5):195, Nov 2019.

\bibitem{owen2013a}
James~E. {Owen} and Yanqin {Wu}.
\newblock {Kepler Planets: A Tale of Evaporation}.
\newblock {\em \apj}, 775(2):105, October 2013.

\bibitem{pepe2014b}
Francesco {Pepe}, David {Ehrenreich}, and Michael~R. {Meyer}.
\newblock {Instrumentation for the detection and characterization of
  exoplanets}.
\newblock {\em \nat}, 513(7518):358--366, September 2014.

\bibitem{peralta2017c}
J.~{Peralta}, R.~{Hueso}, A.~{S{\'a}nchez-Lavega}, Y.~J. {Lee}, A.~Garc{\'\i}a
  {Mu{\~n}oz}, T.~{Kouyama}, H.~{Sagawa}, T.~M. {Sato}, G.~{Piccioni},
  S.~{Tellmann}, T.~{Imamura}, and T.~{Satoh}.
\newblock {Stationary waves and slowly moving features in the night upper
  clouds of Venus}.
\newblock {\em Nature Astronomy}, 1:0187, August 2017.

\bibitem{pollack1993b}
J.~B. {Pollack}, J.~B. {Dalton}, D.~{Grinspoon}, R.~B. {Wattson},
  R.~{Freedman}, D.~{Crisp}, D.~A. {Allen}, B.~{Bezard}, C.~{de Bergh}, L.~P.
  {Giver}, Q.~{Ma}, and R.~{Tipping}.
\newblock {Near-Infrared Light from Venus' Nightside: A Spectroscopic
  Analysis}.
\newblock {\em \icarus}, 103(1):1--42, May 1993.

\bibitem{read2018b}
Peter~L. {Read} and Sebastien {Lebonnois}.
\newblock {Superrotation on Venus, on Titan, and Elsewhere}.
\newblock {\em Annual Review of Earth and Planetary Sciences}, 46:175--202, May
  2018.

\bibitem{ricker2015}
George~R. {Ricker}, Joshua~N. {Winn}, Roland {Vanderspek}, David~W. {Latham},
  G{\'a}sp{\'a}r~{\'A}. {Bakos}, Jacob~L. {Bean}, Zachory~K. {Berta-Thompson},
  Timothy~M. {Brown}, Lars {Buchhave}, Nathaniel~R. {Butler}, R.~Paul {Butler},
  William~J. {Chaplin}, David {Charbonneau}, J{\o}rgen {Christensen-Dalsgaard},
  Mark {Clampin}, Drake {Deming}, John {Doty}, Nathan {De Lee}, Courtney
  {Dressing}, Edward~W. {Dunham}, Michael {Endl}, Francois {Fressin}, Jian
  {Ge}, Thomas {Henning}, Matthew~J. {Holman}, Andrew~W. {Howard}, Shigeru
  {Ida}, Jon~M. {Jenkins}, Garrett {Jernigan}, John~Asher {Johnson}, Lisa
  {Kaltenegger}, Nobuyuki {Kawai}, Hans {Kjeldsen}, Gregory {Laughlin}, Alan~M.
  {Levine}, Douglas {Lin}, Jack~J. {Lissauer}, Phillip {MacQueen}, Geoffrey
  {Marcy}, Peter~R. {McCullough}, Timothy~D. {Morton}, Norio {Narita}, Martin
  {Paegert}, Enric {Palle}, Francesco {Pepe}, Joshua {Pepper}, Andreas
  {Quirrenbach}, Stephen~A. {Rinehart}, Dimitar {Sasselov}, Bun'ei {Sato}, Sara
  {Seager}, Alessandro {Sozzetti}, Keivan~G. {Stassun}, Peter {Sullivan},
  Andrew {Szentgyorgyi}, Guillermo {Torres}, Stephane {Udry}, and Joel
  {Villasenor}.
\newblock {Transiting Exoplanet Survey Satellite (TESS)}.
\newblock {\em Journal of Astronomical Telescopes, Instruments, and Systems},
  1:014003, Jan 2015.

\bibitem{rogers2015a}
Leslie~A. {Rogers}.
\newblock {Most 1.6 Earth-radius Planets are Not Rocky}.
\newblock {\em \apj}, 801(1):41, March 2015.

\bibitem{schubert1980c}
G.~{Schubert}, C.~{Covey}, A.~{del Genio}, L.~S. {Elson}, G.~{Keating},
  A.~{Seiff}, R.~E. {Young}, J.~{Apt}, C.~C. {Counselman}, A.~J. {Kliore},
  S.~S. {Limaye}, H.~E. {Revercomb}, L.~A. {Sromovsky}, V.~E. {Suomi},
  F.~{Taylor}, R.~{Woo}, and U.~{von Zahn}.
\newblock {Structure and circulation of the Venus atmosphere}.
\newblock {\em \jgr}, 85:8007--8025, December 1980.

\bibitem{seager2010}
Sara {Seager} and Drake {Deming}.
\newblock {Exoplanet Atmospheres}.
\newblock {\em \araa}, 48:631--672, September 2010.

\bibitem{sergeev2020}
Denis~E. {Sergeev}, F.~Hugo {Lambert}, Nathan~J. {Mayne}, Ian~A. {Boutle},
  James {Manners}, and Krisztian {Kohary}.
\newblock {Atmospheric Convection Plays a Key Role in the Climate of Tidally
  Locked Terrestrial Exoplanets: Insights from High-resolution Simulations}.
\newblock {\em \apj}, 894(2):84, May 2020.

\bibitem{shields2019a}
Aomawa~L. {Shields}.
\newblock {The Climates of Other Worlds: A Review of the Emerging Field of
  Exoplanet Climatology}.
\newblock {\em \apjs}, 243(2):30, August 2019.

\bibitem{stark2020}
Christopher~C. {Stark}, Courtney {Dressing}, Shannon {Dulz}, Eric {Lopez},
  Mark~S. {Marley}, Peter {Plavchan}, and Johannes {Sahlmann}.
\newblock {Toward Complete Characterization: Prospects for Directly Imaging
  Transiting Exoplanets}.
\newblock {\em \aj}, 159(6):286, June 2020.

\bibitem{stauffer2010b}
John {Stauffer}, Angelle~M. {Tanner}, Geoffrey {Bryden}, Solange {Ramirez},
  Bruce {Berriman}, David~R. {Ciardi}, Stephen~R. {Kane}, Trisha {Mizusawa},
  Alan {Payne}, Peter {Plavchan}, Kaspar {von Braun}, Pamela {Wyatt}, and
  J.~Davy {Kirkpatrick}.
\newblock {Accurate Coordinates and 2MASS Cross Identifications for (Almost)
  All Gliese Catalog Star}.
\newblock {\em \pasp}, 122(894):885, August 2010.

\bibitem{takagi2007}
M.~{Takagi} and Y.~{Matsuda}.
\newblock {Effects of thermal tides on the Venus atmospheric superrotation}.
\newblock {\em Journal of Geophysical Research (Atmospheres)}, 112(D9):D09112,
  May 2007.

\bibitem{taylor2009}
F.~{Taylor} and D.~{Grinspoon}.
\newblock {Climate evolution of Venus}.
\newblock {\em Journal of Geophysical Research (Planets)}, 114(E11):E00B40, May
  2009.

\bibitem{taylor2018}
Fredric~W. {Taylor}, H{\r{a}}kan {Svedhem}, and James~W. {Head}.
\newblock {Venus: The Atmosphere, Climate, Surface, Interior and Near-Space
  Environment of an Earth-Like Planet}.
\newblock {\em \ssr}, 214(1):35, Feb 2018.

\bibitem{reportluvoir}
{The LUVOIR Team}.
\newblock {The LUVOIR Mission Concept Study Final Report}.
\newblock {\em arXiv e-prints}, page arXiv:1912.06219, December 2019.

\bibitem{turbet2021}
Martin {Turbet}, Emeline {Bolmont}, Guillaume {Chaverot}, David {Ehrenreich},
  J{\'e}r{\'e}my {Leconte}, and Emmanuel {Marcq}.
\newblock {Day-night cloud asymmetry prevents early oceans on Venus but not on
  Earth}.
\newblock {\em \nat}, 598(7880):276--280, October 2021.

\bibitem{turbet2016}
Martin {Turbet}, J{\'e}r{\'e}my {Leconte}, Franck {Selsis}, Emeline {Bolmont},
  Fran{\c{c}}ois {Forget}, Ignasi {Ribas}, Sean~N. {Raymond}, and Guillem
  {Anglada-Escud{\'e}}.
\newblock {The habitability of Proxima Centauri b. II. Possible climates and
  observability}.
\newblock {\em \aap}, 596:A112, December 2016.

\bibitem{way2017b}
M.~J. {Way}, I.~{Aleinov}, David~S. {Amundsen}, M.~A. {Chand ler}, T.~L.
  {Clune}, A.~D. {Del Genio}, Y.~{Fujii}, M.~{Kelley}, N.~Y. {Kiang},
  L.~{Sohl}, and K.~{Tsigaridis}.
\newblock {Resolving Orbital and Climate Keys of Earth and Extraterrestrial
  Environments with Dynamics (ROCKE-3D) 1.0: A General Circulation Model for
  Simulating the Climates of Rocky Planets}.
\newblock {\em \apjs}, 231(1):12, Jul 2017.

\bibitem{way2020}
M.~J. {Way} and Anthony~D. {Del Genio}.
\newblock {Venusian Habitable Climate Scenarios: Modeling Venus Through Time
  and Applications to Slowly Rotating Venus-Like Exoplanets}.
\newblock {\em Journal of Geophysical Research (Planets)}, 125(5):e06276, May
  2020.

\bibitem{way2016}
M.~J. {Way}, Anthony~D. {Del Genio}, Nancy~Y. {Kiang}, Linda~E. {Sohl},
  David~H. {Grinspoon}, Igor {Aleinov}, Maxwell {Kelley}, and Thomas {Clune}.
\newblock {Was Venus the first habitable world of our solar system?}
\newblock {\em \grl}, 43(16):8376--8383, Aug 2016.

\bibitem{winn2015}
J.~N. {Winn} and D.~C. {Fabrycky}.
\newblock {The Occurrence and Architecture of Exoplanetary Systems}.
\newblock {\em \araa}, 53:409--447, August 2015.

\bibitem{wolfgang2016}
Angie {Wolfgang}, Leslie~A. {Rogers}, and Eric~B. {Ford}.
\newblock {Probabilistic Mass-Radius Relationship for Sub-Neptune-Sized
  Planets}.
\newblock {\em \apj}, 825(1):19, July 2016.

\bibitem{wordsworth2015a}
Robin {Wordsworth}.
\newblock {Atmospheric Heat Redistribution and Collapse on Tidally Locked Rocky
  Planets}.
\newblock {\em \apj}, 806(2):180, June 2015.

\bibitem{yang2014b}
Jun {Yang}, Gwena{\"e}l {Bou{\'e}}, Daniel~C. {Fabrycky}, and Dorian~S.
  {Abbot}.
\newblock {Strong Dependence of the Inner Edge of the Habitable Zone on
  Planetary Rotation Rate}.
\newblock {\em \apjl}, 787(1):L2, May 2014.

\bibitem{yang2013}
Jun {Yang}, Nicolas~B. {Cowan}, and Dorian~S. {Abbot}.
\newblock {Stabilizing Cloud Feedback Dramatically Expands the Habitable Zone
  of Tidally Locked Planets}.
\newblock {\em \apjl}, 771(2):L45, July 2013.

\bibitem{yang2019b}
Jun {Yang}, J{\'e}r{\'e}my {Leconte}, Eric~T. {Wolf}, Timothy {Merlis}, Daniel
  D.~B. {Koll}, Fran{\c{c}}ois {Forget}, and Dorian~S. {Abbot}.
\newblock {Simulations of Water Vapor and Clouds on Rapidly Rotating and
  Tidally Locked Planets: A 3D Model Intercomparison}.
\newblock {\em \apj}, 875(1):46, April 2019.

\bibitem{yang2014c}
Jun {Yang}, Yonggang {Liu}, Yongyun {Hu}, and Dorian~S. {Abbot}.
\newblock {Water Trapping on Tidally Locked Terrestrial Planets Requires
  Special Conditions}.
\newblock {\em \apjl}, 796(2):L22, December 2014.

\bibitem{young1987b}
R.~E. {Young}, R.~L. {Walterscheid}, G.~{Schubert}, A.~{Seiff}, V.~M. {Linkin},
  and A.~N. {Lipatov}.
\newblock {Characteristics of gravity waves generated by surface topography on
  Venus: comparison with the Vega balloon results.}
\newblock {\em Journal of Atmospheric Sciences}, 44:2628--2639, September 1987.

\bibitem{young1994c}
Richard~E. {Young}, Richard~L. {Walterscheid}, Gerald {Schubert}, Leonhard
  {Pfister}, Howard {Houben}, and Duane~L. {Bindschadler}.
\newblock {Characteristics of Finite Amplitude Stationary Gravity Waves in the
  Atmosphere of Venus.}
\newblock {\em Journal of Atmospheric Sciences}, 51(13):1857--1875, July 1994.

\bibitem{zhang2009}
T.~L. {Zhang}, J.~{Du}, Y.~J. {Ma}, H.~{Lammer}, W.~{Baumjohann}, C.~{Wang},
  and C.~T. {Russell}.
\newblock {Disappearing induced magnetosphere at Venus: Implications for
  close-in exoplanets}.
\newblock {\em \grl}, 36(20):L20203, October 2009.

\end{thebibliography}


\end{document}